\documentclass[10pt]{article}

\usepackage[margin=1.5in]{geometry}
\usepackage{amsmath,amsfonts}
\usepackage{xcolor}
\usepackage{graphicx}
\usepackage{subfig}
\usepackage{nicefrac}
\usepackage[hidelinks]{hyperref}
\usepackage{cleveref}

\usepackage[binary-units]{siunitx}
\usepackage{caption}
\usepackage{tabularx,multirow,booktabs}
\newcolumntype{L}{>{\raggedright\arraybackslash}X}
\newcolumntype{C}{>{\centering\arraybackslash}X}
\newcolumntype{R}{>{\raggedleft\arraybackslash}X}

\begin{document}

\title{\bf Enhancing Scalability \\ of a Matrix-Free Eigensolver \\ for Studying Many-Body Localization}

\author{%
  Roel~Van~Beeumen\textsuperscript{1,}%
    \footnote{Lawrence Berkeley National Laboratory,
              1 Cyclotron Road, Berkeley, CA 94720.
              Email: \texttt{rvanbeeumen@lbl.gov}.} ,
  Khaled~Z.~Ibrahim\textsuperscript{1},\\
  Gregory~D.~Kahanamoku--Meyer\textsuperscript{2},
  Norman~Y.~Yao\textsuperscript{2}, and
  Chao~Yang\textsuperscript{1,}}
\date{\small%
\textsuperscript{1}Computational Research Division,
Lawrence Berkeley National Laboratory, Berkeley, CA \\
\textsuperscript{2}Department of Physics,
University of California at Berkeley, Berkeley, CA
}

\maketitle

\begin{abstract}
In [Van Beeumen, et. al, HPC Asia 2020,
\url{https://www.doi.org/10.1145/3368474.3368497}] a scalable and matrix-free
eigensolver was proposed for studying the many-body localization (MBL)
transition of two-level quantum spin chain models with nearest-neighbor $XX+YY$
interactions plus $Z$ terms.
This type of problem is computationally challenging because the vector space
dimension grows exponentially with the physical system size, and averaging over
different configurations of the random disorder is needed to obtain relevant
statistical behavior.
For each eigenvalue problem, eigenvalues from different regions of the spectrum
and their corresponding eigenvectors need to be computed.
Traditionally, the interior eigenstates for a single eigenvalue problem are
computed via the shift-and-invert Lanczos algorithm.
Due to the extremely high memory footprint of the LU factorizations, this
technique is not well suited for large number of spins $L$, e.g., one needs
thousands of compute nodes on modern high performance computing infrastructures
to go beyond $L = 24$.
The matrix-free approach does not suffer from this memory bottleneck, however,
its scalability is limited by a computation and communication imbalance.
We present a few strategies to reduce this imbalance and to significantly
enhance the scalability of the matrix-free eigensolver.
To optimize the communication performance, we leverage the consistent space
runtime, CSPACER, and show its efficiency in accelerating the MBL irregular
communication patterns at scale compared to optimized MPI non-blocking two-sided
and one-sided RMA implementation variants.
The efficiency and effectiveness of the proposed algorithm is demonstrated by
computing eigenstates on a massively parallel many-core high performance computer.
\end{abstract}

\paragraph{Keywords:}
Quantum many-body problem, many-body localization, eigenvalue problem,
matrix-free eigensolver, LOBPCG method, preconditioner, scalability,
graph partitioning, METIS, communication optimization, CSPACER.

\section{Introduction}
\label{sec:intro}

A fundamental assumption in the traditional theory of statistical mechanics is that an isolated system will in general reach an equilibrium state, or \emph{thermalize}.
As early as the mid-20\textsuperscript{th} century, Anderson demonstrated that a single particle moving in a highly disordered landscape can violate this assumption  \cite{ande1958}.
While surprising, that result does not readily extend to many-particle systems that exhibit strong interactions between the constitutent particles.
The question of whether a similar effect could manifest in a strongly-interacting many-body system remained open for decades.
This elusive phenomenon has been termed ``many-body localization'' (MBL).

Recently, advances in both high performance computing and experimental control of individual quantum particles have begun to yield insight into MBL.
Both experimental \cite{scho2015,chhi2016,smle2016,kosc2019,bolu2017,luri2019} and numerical \cite{lula2015,jona2015,cufe2012,bana2013,pahu2010} results have shown evidence of localization in small strongly-interacting multiparticle systems of 10-20 spins.
Unfortunately, extrapolating results from these small system sizes to the infinitely-large thermodynamic limit has proven difficult.
This lack of clarity has inspired a vigorous debate in the community about precisely what can be learned from small-size results.
For example, it has been proposed that certain features do not actually exist at infinite system size \cite{drhu2016}, and even that MBL itself is only a finite-size effect \cite{subo2019,abba2019}!

The primary goal of most studies is to identify and characterize a \emph{localization transition}.
In the thermodynamic limit, as the strength of the system's disorder increases, theory predicts a sharp, sudden change from a thermal to a localized state.
Unfortunately, in the small systems available for study, that sharp transition turns into a smooth \emph{crossover}, leading to the confusion about what constitutes the transition itself.
Numerical evidence suggests that the transition sharpens rapidly as system size increases, so accessing as large systems as possible is imperative for investigating MBL.

In pursuit of that goal, Luitz et al. used large-scale numerical linear algebra to show a localization transition for system sizes up to $L = 22$ \cite{lula2015}, and in a following paper extracted useful data up to $L = 24$ \cite{pima2018}.
In order to compute interior eigenstates for the MBL problem, the shift-and-invert Lanczos algorithm was used in combination with sparse direct solvers for solving the linear systems.
One of the major disadvantages of this technique is that constructing the LU factorizations becomes extremely memory demanding, due to the so called fill in, for large number of spins $L$.
\Cref{tab:memory} shows that the memory footprint of the LU factorization computed via STRUMPACK \cite{ghli2017} grows rapidly as function of $L$.
See also \cite{pima2018}.
Hence, thousands of nodes on modern high performance computing infrastructures are needed to go beyond $L = 24$.

\begin{table}[hbtp]
\small\centering
\caption{%
Total memory footprint as a function of the spin chain length $L$ for LU
factorizations, computed via STRUMPACK, and the matrix-free LOBPCG algorithm
\cite{vbkm2020}, with block size 64. The problem size is given by $n$.%
\label{tab:memory}}
\begin{tabularx}{3in}{C|S[table-format = 8]|S[table-format = 3,table-space-text-post = \,\si{\mega\byte}]|S[table-format = 3,table-space-text-post = \,\si{\mega\byte}]}
\toprule
$L$ & \multicolumn{1}{c|}{$n$} & \multicolumn{1}{c|}{STRUMPACK} & \multicolumn{1}{c}{LOBPCG(64)} \\
\midrule
16  &     12870  &   66\,\si{\mega\byte}  &    8\,\si{\mega\byte} \\
18  &     48620  &  691\,\si{\mega\byte}  &   31\,\si{\mega\byte} \\
20  &    184756  &    8\,\si{\giga\byte}  &  118\,\si{\mega\byte} \\
22  &    705432  &   92\,\si{\giga\byte}  &  451\,\si{\mega\byte} \\
24  &   2704156  &    1\,\si{\tera\byte}  &    2\,\si{\giga\byte} \\
26  &  10400600  &   15\,\si{\tera\byte}  &    7\,\si{\giga\byte} \\
\bottomrule
\end{tabularx}
\end{table}

To overcome the memory bottleneck that the shift-and-invert Lanczos algorithm
faces, we recently proposed in \cite{vbkm2020} a matrix-free locally optimal
block preconditioned conjugate gradient (LOBPCG) algorithm.
As shown in \Cref{tab:memory}, this approach reduces the memory footprint by
several orders of magnitude, e.g., from 15\,TB to only 7\,GB for $L = 26$, and
enables simulating spin chains on even a single node, up to $L = 24$.
In contrast to the shift-and-invert Lanczos algorithm, where the dominant
computational cost is the construction of the LU factorization, the dominant
computational cost of the LOBPCG algorithm is the (block) matrix-vector
(MATVEC) product.
As illustrated in \cite{vbkm2020}, the scalability of this MATVEC is limited at
high concurrency which is due to a computation and communication imbalance.
In the current paper, we present different strategies to overcome this imbalance
and to significantly enhance the scalability of the matrix-free eigensolver.

The paper is organized as follows.
We first review the Heisenberg spin model and MBL metrics in \Cref{sec:problem}.
The multiple levels of concurrency and the matrix-free LOBPCG eigensolver are
discussed in \Cref{sec:simul}.
Next, we present the balancing of computation and communication within the MATVEC
in \Cref{sec:bal} and the optimization of the communication performance in
\Cref{sec:comm}.
Then in \Cref{sec:exp}, we illustrate the different proposed strategies for
improving the computation and communication imbalance of the matrix-free LOBPCG
eigensolver for $L = 24$ and $L = 26$ problems.
Finally, the main conclusions are formulated in \Cref{sec:concl}.

\section{Problem Formulation}
\label{sec:problem}

In this section we briefly review the properties of the spin chain model that most frequently is studied by numerical simulations of MBL.

\subsection{Heisenberg Spin Model}
\label{sec:model}

We consider the nearest-neighbor interacting Heisenberg spin model with random on-site fields:
\begin{equation}
H = \sum_{\langle i,j \rangle} \vec{S}_i \cdot \vec{S}_j + \sum_i h_i S_i^z,
\label{eq:model}
\end{equation}
where the angle brackets denote nearest-neighbor $i$ and $j$, $h_i$ is sampled from a uniform distribution $[-w,w]$ with $w \in \mathbb{R}_0^+$, and
\begin{displaymath}
\vec{S}_i \cdot \vec{S}_j = S_i^x S_j^x + S_i^y S_j^y + S_i^z S_j^z,
\end{displaymath}
where $S_i^\alpha = \frac{1}{2} \sigma_i^\alpha$, with $\sigma_i^\alpha$ the Pauli matrices operating on lattice site $i$ and $\alpha \in \{x,y,z\}$.
The parameter $w$ is called the \emph{disorder strength}, and is responsible for inducing the MBL transition.
The values $h_i$ are sampled randomly each time the Hamiltonian is instantiated, and the relevant physics lies in the statistical behavior of the set of all such Hamiltonians.
The individual Hamiltonians $H$ with independently sampled $h_i$ are called \emph{disorder realizations}.

Note that in \eqref{eq:model} each term of each sum has an implied tensor product with the identity on all the sites not explicitly written.
Consequently, the Hamiltonian for $L$ spins is a symmetric matrix of dimension $N = 2^L$ and exhibits the following tensor product structure
\begin{equation}
H = \sum_{i=1}^{L-1} I \otimes \cdots \otimes I \otimes H_{i,i+1} \otimes I \otimes \cdots \otimes I 
  + \sum_{i=1}^L I \otimes \cdots \otimes I \otimes h_i S_i^z \otimes I \otimes \cdots \otimes I,
\label{eq:Hmatrix}
\end{equation}
where $H_{i,i+1} = S_i^x S_{i+1}^x + S_i^y S_{i+1}^y + S_i^z S_{i+1}^z$ is a 4-by-4 real matrix and $I$ is the 2-by-2 identity matrix.
Remark that by definition, all matrices $H_{i,i+1}$ are the same and independent of the site $i$.
For our experiments, we use open boundary conditions, meaning that the nearest-neighbor terms do not wrap around at the end of the spin chain.
Open boundary conditions can be considered to yield a larger \emph{effective} system size because of the reduced connectivity.

The state of each spin is described by a vector in $\mathbb{C}^2$, and the configuration of the entire $L$-spin system can be described by a vector on the tensor product space $(\mathbb{C}^2)^{\otimes L}$.
In this specific case, however, the Hamiltonian's matrix elements happen to all be real, so we do not include an imaginary part in any of our computations.
Furthermore, our Hamiltonian commutes with the total magnetization in the $z$ direction, $S^z = \sum_{i=1}^L S_i^z$.
Thus it can be block-diagonalized in sectors characterized by $S^z \in [-\nicefrac{L}{2},-\nicefrac{L}{2}+1,\ldots,\nicefrac{L}{2}-1,\nicefrac{L}{2}]$.
The vector space corresponding to each sector has dimension $n = {L \choose S^z + \nicefrac{L}{2}}$ such that the largest sector's dimension is $n=\frac{L!}{(\nicefrac{L}{2})!(\nicefrac{L}{2})!}$, and this corresponds to the actual dimension of the matrices on which we operate, see \Cref{tab:memory}.
While these subspaces are smaller than the full space, their size still grows exponentially with the number of spins $L$.
Thus, the problem becomes difficult rapidly as $L$ increases.
Furthermore, the density of eigenvalues in the middle of the spectrum increases exponentially with $L$. Thus the tolerance used to solve for these internal eigenvalues must be made tighter rapidly as $L$ increases.

\subsection{Many-Body Localization}
\label{sec:mbl}

With the problem's matrix clearly defined, we now review ways for
quantifying localization from the eigenvalues and eigenvectors.
There are multiple quantities that can be used for identifying localization.

One of the commonly used quantities is the \emph{adjacent gap ratio}
\cite{oghu2007,subo2019,jona2015,cufe2012}.
This approach is based on the statistical distribution of the eigenvalues of
different disorder realizations, hence, only eigenvalues need to be computed.
Random matrix theory informs us that the statistical distribution of eigenvalues
will differ between localizing and thermalizing Hamiltonians \cite{oghu2007}.
In particular, we expect eigenvalues of a thermal Hamiltonian to \emph{repel}
each other, i.e., hybridization of eigenvectors prevents them from generally
coming too close to one another.
The eigenvalues of a localized Hamiltonian should not display this behavior:
we expect them to be Poisson distributed.
Therefore, we can measure localization by comparing the relative size of gaps
between the eigenvalues.
Thermal Hamiltonians will generally have more consistently sized gaps due to
level repulsion.
However, this technique suffers from large statistical noise and thus requires
many samples to be usable.

Another quantity for measuring localization is the \emph{eigenstate
entanglement entropy} \cite{vbkm2020} which is based on the eigenvectors of
the Hamiltonians.
In a thermal system, we expect quantum entanglement to be widespread, while
in a localized system, the entanglement is not expected to be extensive.
This idea can be quantified by choosing a \emph{cut} which divides the spin
chain into two pieces, and measuring the entanglement across it.
Not only the value of the entropy changes during the localization transition:
the statistics change as well.
When compared across disorder realizations, the thermal entanglement entropy
has small variance.
During the transition, however, the entanglement entropy depends strongly on the
specific disorder realization and thus the statistic will have a large variance.
Empirically, examining the \emph{variance} of the entanglement entropy is one of
the best ways to identify the localization transition and requires fewer samples
than the adjacent gap ratio approach.

\section{Massively Parallel Simulation}
\label{sec:simul}

In order to maximally reduce the finite size effects on the determination of
the MBL transition point, we need to study spin models with as many spins as
possible.
Consequently, this problem is computationally demanding and requires lots of
resources.

\subsection{Multiple Levels of Concurrency}
\label{sec:concurrency}

The MBL study allows for at least 4 levels of concurrency.
The first level corresponds to the need of averaging over (many) different
and independently sampled \emph{disorder realizations} in order to obtain
relevant statistical behavior.
Since the \emph{disorder strength} is responsible for inducing the MBL
transition, we also have to vary the disorder strength, giving rise to
the second level of concurrency.
The third level corresponds to the \emph{eigenvalue chunks}, i.e., for
each (large) eigenvalue problem, originating from one disorder realization
and a particular disorder strength, we have to compute eigenvalues from
different regions of the spectrum and their corresponding eigenvectors.

All previous levels of concurrency are completely independent and can be
implemented in a massively parallel fashion by making use of iterative
eigensolvers.
In this paper, we therefore only focus on the forth level of parallelism
taking place within these eigensolvers.
Although most iterative eigensolvers follow a rather sequential procedure,
each of the different steps within one iteration can be implemented in parallel.

\subsection{Matrix-Free LOBPCG Eigensolver}
\label{sec:solver}

The Locally Optimal Block Preconditioned Conjugate Gradient (LOBPCG)
algorithm \cite{knya2001,dush2018} is a widely used eigensolver for
computing the smallest or largest eigenvalues and corresponding
eigenvectors of large-scale symmetric matrices.
Key features of the LOBPCG algorithm are:
  (i) It is matrix-free, i.e., the solver does not require storing the
      coefficient matrix explicitly and it access the matrix by only
      evaluating matrix-vector products;
 (ii) It is a block method, which allows for efficient matrix-matrix
      operations on modern computing architectures;
(iii) It can take advantage of preconditioning, in contrast to,
      for example, the Lanczos algorithm.

The standard LOBPCG algorithm allows for computing either the lower or
upper part of the spectrum.
In order to compute interior eigenvalues and their corresponding eigenvectors,
we make use of the so called \emph{spectral fold} transformation \cite{wazu1994}
\begin{displaymath}
(H - \sigma I)^2,
\end{displaymath}
where $\sigma \in\mathbb{R}$ is the shift around which we want to compute
eigenvalues.
This spectral transformation maps all eigenvalues to the positive real axis
and the ones closest to the shift $\sigma$ to the lower edge close to 0.
Hence, after applying this transformation, we can also use the LOBPCG
eigensolver for computing interior eigenvalues.
Because the transformed eigenvalue problem
\begin{displaymath}
(H - \sigma I)^2 x = \lambda x
\end{displaymath}
is symmetric positive definite, we use a diagonal (Jacobi) preconditioned
conjugate gradient (PCG) method as preconditioner for the LOBPCG eigensolver.
For more details on the matrix-free LOBPCG eigensolver used in the particular
case of studying many-body localization, we refer to \cite{vbkm2020}.

In contrast to the shift-and-invert Lanczos algorithm, where the dominant
computational cost is the construction of the LU factorization, the dominant
computational cost of the LOBPCG and PCG algorithms is the (block) MATVEC.
In the remainder of the paper, we therefore will mainly focus on enhancing the
scalability of the MATVEC.

\section{Balancing Computation and Communication}
\label{sec:bal}

In this section we have a closer look at the MBL (block) matrix-vector product
and focus on how to enhance its scalability by reducing the computation and
communication imbalance.

\subsection{Matrix-Free Matrix-Vector Product}

As a starting point, we take the hybrid MPI--OpenMP MATVEC introduced in
\cite{vbkm2020}.
This matrix-free MATVEC uses one MPI rank per node and OpenMP for on-node parallelism.
The block of vectors to be multiplied by the Hamiltonian matrix is partitioned
by rows and distributed among different MPI ranks (and nodes).
Within each MPI rank, a local sparse MATVEC is performed. A subset of the  rows in the local vector block need to be sent to other MPI ranks to be multiplied and accumulated on the target MPI ranks. Each MPI rank also receives vector block contributions from other MPI ranks to be combined with the local product. 
It has been illustrated in \cite{vbkm2020} that the parallel MATVEC implementation using
non-blocking MPI communication, in combination with overlapping communication
and local computation, results in the best performance.

\begin{figure}[b!]
\centering
\includegraphics{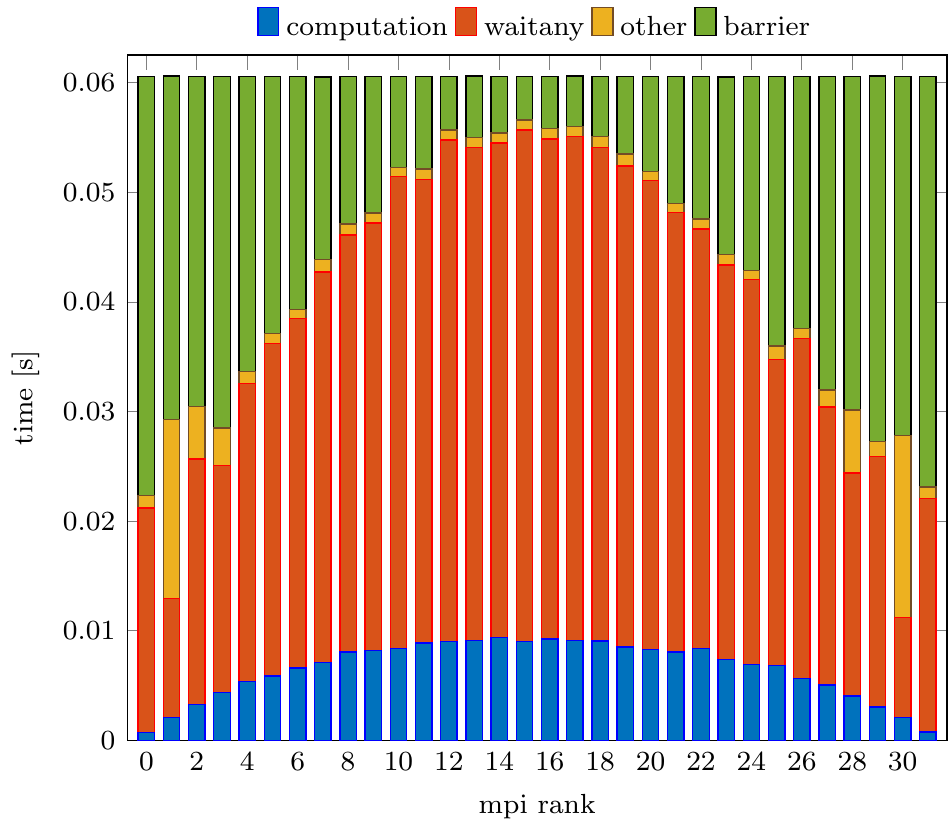}
\caption{\label{fig:matvec-unbalance}%
Average wall clock time as a function of the MPI rank for the different
components of the $L = 26$ non-blocking MPI--OpenMP MATVEC with block size 64.}
\end{figure}

\Cref{fig:matvec-unbalance} shows the average wall clock time as a function of
the rank for 100 samples of the $L = 26$ non-blocking MPI-OpenMP MATVEC.
The different amounts of time spend in the MPI barrier shows clearly the high
computation and communication imbalance of this MATVEC.
Note that the local computation time can be fully overlapped with the
communication and that the computation time shown only corresponds to the
\emph{remote} computation time which can only start once the incoming data
has arrived.
\Cref{fig:matvec-unbalance} also illustrates that this MATVEC is communication
dominated.
The difference between computation and communication time further increases for
higher concurrency.
Therefore, in the remainder of this paper, we will focus on different strategies
for reducing and optimizing the communication time.

\subsection{Graph Partitioning}
The row partition of the vectors among different MPI ranks corresponds to a partition of the adjacency graph induced by the many-body Hamiltonian, with vertices mapped rows or columns, and edges mapped to nonzero elements of the Hamiltonian. Within each partition, vertices that are not connected with vertices in other partitions by edges are strictly local. The corresponding rows do not need to be sent to other MPI ranks. The vertices that are connected to vertices in other partitions are called \textit{shared rows}. They need to be sent to other MPI ranks in a parallel MATVEC implementation.

The state ordering for the hybrid MPI--OpenMP MATVEC in \cite{vbkm2020} leads to
a simple communication pattern where, up to $\sim$50 ranks, only communication
with neighboring ranks in a linear topology is required.
The ordering of the states also allows for efficiently computing the
off-diagonal element indices on the fly for applying the matrix-free MATVEC.
However, as shown in \Cref{fig:matvec-unbalance}, the communication is largely
imbalanced and far from optimal.
In particular, the graph for the Hamiltonian is quite nonuniform, and vertices
in the \emph{middle} are much more densely interconnected than vertices on the
\emph{edge}.

One of the properties of the Heisenberg spin model with random on-site fields is
that the sparsity pattern of \eqref{eq:Hmatrix} does not change for different disorder realizations, neither does the corresponding matrix graph.
Therefore, we can apply graph partition techniques in order to reduce the
communication volume and better balance the communication time among the
different MPI ranks.
A comparison of the communication volume, as a function of the total number of
MPI ranks, between the hybrid MPI--OpenMP state ordering used in \cite{vbkm2020}
and the METIS $k$-way graph partitioning \cite{kaku1998} reordering of the
states is shown in \Cref{tab:comm}.
For the METIS graph partitioning we used the objective function for total
communication volume minimization \cite{kary2013}.

\begin{table}[hbtp]
\small\centering
\caption{%
Communication volume as a function of the total number of MPI ranks for the
$L = 26$ MPI--OpenMP MATVEC with the state ordering in \cite{vbkm2020} and
the METIS reordering.\label{tab:comm}}
\subfloat[Total communication volume\label{tab:comm-tot}]{%
\begin{tabularx}{3.35in}{C|S[table-format = 7]|S[table-format = 7]|S[table-format = 7]|S[table-format = 7]}
\toprule
ranks  & \multicolumn{1}{c|}{4} &
         \multicolumn{1}{c|}{8} &
         \multicolumn{1}{c|}{16} &
         \multicolumn{1}{c}{32} \\
\midrule
\cite{vbkm2020}
       &  1097358  &  2552781  &  4788256  &  9620799 \\
METIS  &  1406736  &  2350508  &  4082179  &  5893137 \\
\midrule
gain   & \multicolumn{1}{c|}{$-$28\,\%}
       & \multicolumn{1}{c|}{\bf 8\,\%}
       & \multicolumn{1}{c|}{\bf 15\,\%}
       & \multicolumn{1}{c}{\bf 39\,\%} \\
\bottomrule
\end{tabularx}}\\
\subfloat[Maximum communication volume per rank\label{tab:comm-max}]{%
\begin{tabularx}{3.35in}{C|S[table-format = 7]|S[table-format = 7]|S[table-format = 7]|S[table-format = 7]}
\toprule
ranks  & \multicolumn{1}{c|}{4} &
         \multicolumn{1}{c|}{8} &
         \multicolumn{1}{c|}{16} &
         \multicolumn{1}{c}{32} \\
\midrule
\cite{vbkm2020}
       &  376007  &  408962  &  416484  &  419402 \\
METIS  &  481086  &  389561  &  348559  &  277669 \\
\midrule
gain   & \multicolumn{1}{c|}{$-$28\,\%}
       & \multicolumn{1}{c|}{\bf 5\,\%}
       & \multicolumn{1}{c|}{\bf 16\,\%}
       & \multicolumn{1}{c}{\bf 34\,\%} \\
\bottomrule
\end{tabularx}}
\end{table}

\Cref{tab:comm} shows that for the $L = 24$ MATVEC beyond 4 MPI ranks both the
total communication volume as well as the maximum communication volume per rank
can significantly be reduced by using the METIS reordering.
Note that this reordering will have an effect on both the computation and
communication within the MATVEC.
First, although the communication volume is reduced, the METIS reordering of
the states results in a more complicated communication pattern compared to the
original MPI--OpenMP one in \cite{vbkm2020}.
Hence, each rank needs, in general, to communicate with more than 2 other ranks.
Second, the METIS reordering of the states is also less structured and therefore
the remote MATVEC computation will require more complicated lookup tables.
However, since the MATVEC is communication dominant and we maximally overlap
computation with communication, the extra overhead from a slightly slower
computational portion will be marginal.

\section{Communication Performance Optimization}
\label{sec:comm}

The load balancing of computation using METIS affects the communication pattern---specifically,  the number of messages per rank from being constant to being a function of the rank count.

\begin{figure}[hbtp]
\centering
\includegraphics[width=0.7\textwidth]{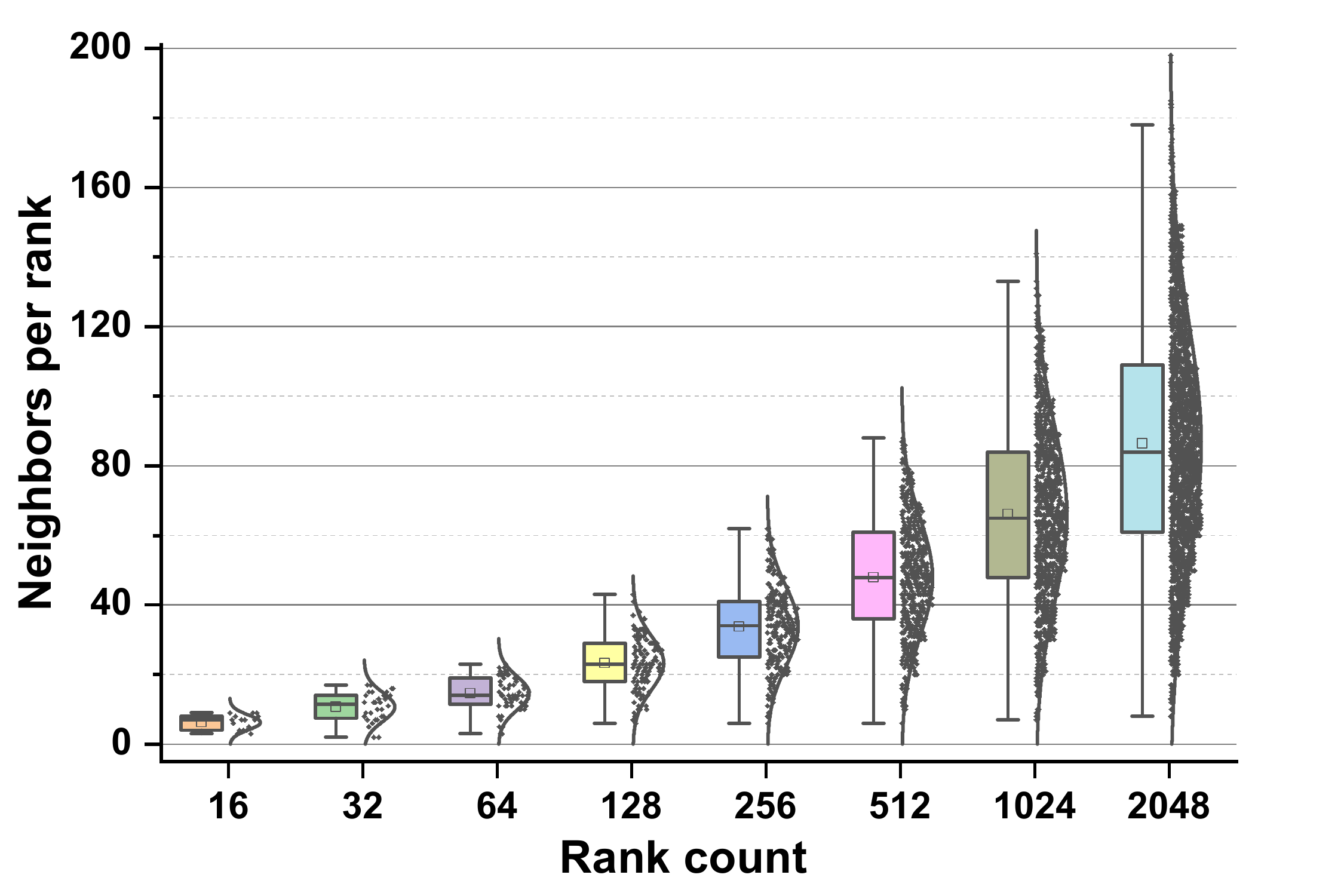}
\caption{\label{fig:L26msgs}%
The distribution of neighbor count for vector data exchange as we scale the job
size due to load balancing the $L=26$ MATVEC.
Not only the number of communication messages increases, but also the variability
increases significantly while scaling.}
\end{figure}

\Cref{fig:L26msgs} shows the distribution of the neighbor count a rank needs to
communicate the vector with, as we increase the job size.
With METIS partitioning, depending on the rank position within a job, a different
number of neighbors are involved in the vector exchange.
With each of these neighbors, a rank needs to communicate a portion of their
assigned vector.
As shown in \Cref{fig:L26msgs}, for the $L=26$ problem, the number of neighbors
increases super-linearly with respect to $\log(r)$, where $r$ is the rank count.

\begin{figure}[hbtp]
\centering
\includegraphics[width=0.7\textwidth]{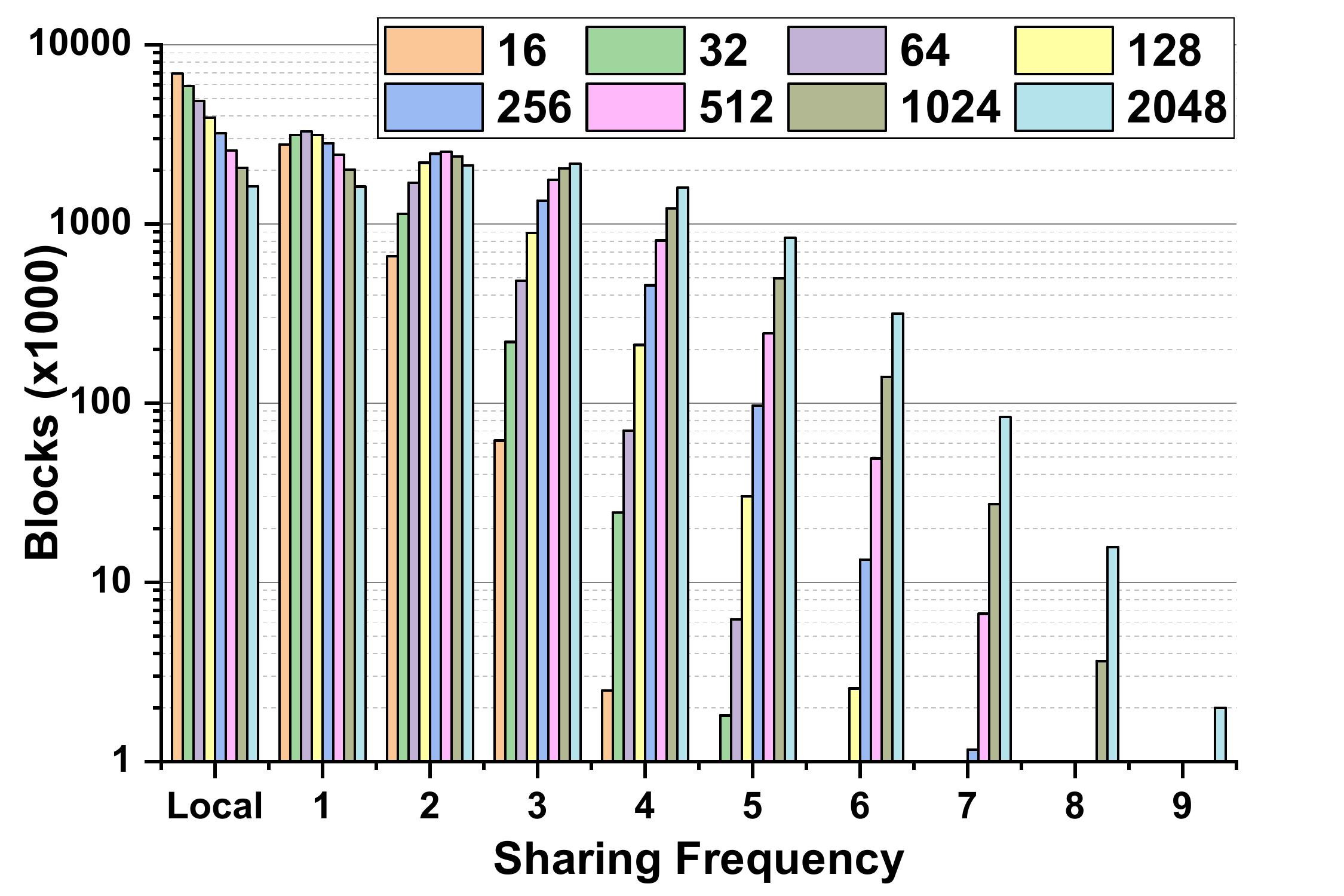}
\caption{\label{fig:BlockSharing}%
The amount of data to be communicate (or ``shared'') among different ranks in
the $L=26$ MATVEC communication phase as we scale the computation.
The sharing level increases as the number of ranks increases,
making the scaling communication bound.}
\end{figure}

\Cref{fig:BlockSharing} shows the number of rows, of the partitioned block
vector, each rank exchanges with their neighbors as a function of the
\emph{sharing level}, i.e., the number of neighbors a row block is sent to.
From this figure, we see that the number of row blocks not involved in data
exchange decreases as we strong scale the computation, hence, making it
necessary to communicate an increasing fraction of rows.
Moreover, the sharing level increase as well, making it necessary to exchange
the same row with multiple ranks.
Such sharing makes the volume of communicated data to decrease sub-linearly with
the rank count.
Overall, the total volume of communicated data increases as we scale the job,
roughly proportional to the $\sqrt{r}$.

Fortunately, the sparsity pattern that influences the communication pattern
remains unchanged across iterations.
As such, one could classify this communication pattern as a static irregular
one, and we can construct all needed information about the communication pattern
before the communication starts.
Although the algorithm relies on other communication primitives, such as
\texttt{allreduce}, they do not significantly contribute to the execution time.

The discussed attributes show the challenge in strong-scaling the computation, which could be summarized by the need for processing an increasingly large number of small messages.

\subsection{Structuring Communication using MPI}
MPI provides multiple techniques to implement a communication pattern. While not claiming that we exhausted all possible methods, we explored a few widely used techniques that are likely to serve our communication pattern.

The most natural way to implement our matrix-vector communication pattern is to use non-blocking point-to-point transfers. We overlapped local computation with some of the communication cost by assigning a thread to progress communication in the background while performing local computations.

In addition to using point-to-point communication primitives, we explored the use of MPI-3 non-blocking \texttt{alltoallv} collective primitives~\cite{hiwe2016}, and MPI-3 RMA~\cite{hodi2015}. The collective-based implementation matches more or less the non-blocking point-to-point and is omitted for brevity. We found the MPI one-sided implementation efficient at small scale, but the scaling behavior is inefficient in our experience on the Cray XC40 system.

\subsection{Structuring Communication using CSPACER}

In this study, we explored the use of CSPACER~\cite{CSPACER_XC}, Consistent SPACE Runtime, which provides a low overhead communication abstraction for irregular communication patterns. The runtime extends the support of the consistency space abstraction~\cite{ibra2019} to Cray systems. The runtime could interoperate with the MPI runtime, allowing for incremental integration and tackling communication hotspots while retaining the bulk of MPI's communication code.

The space consistency abstraction~\cite{ibra2019} is a generalization of full-empty synchronization for distributed computing, where each memory region is associated with a counter that determines its consistency.
A memory space becomes consistent, i.e., ready for consumption, when the counter matches a specific consistency tag. To construct a consistent state, a space typically receives one or multiple transfers from one or more producers. The runtime provides APIs to facilitate checking the consistency of a space for consumers, but it does not provide the functionality of tracking the completions for individual data transfers. 
The runtime relies on symmetric space allocations across a team of ranks, and supports communication primitives such as one-sided put, various collectives that are implemented as patterns of multiple underlying primitives. Due to its simple design, the CSPACER runtime enjoys a low injection overhead, in the range of 0.4\,\si{\micro\second} on KNL architectures, and provides good scalability, especially for irregular communication patterns.

The space consistency adopts a memory-centric approach while orchestrating communication across ranks, rather than relying on transfer-centric strategies.  It supports multiple mechanisms for issuing transfer operations that help achieving a consistent state, including concurrent threaded injection from an OpenMP parallel region, pipelined injection and progress, etc. The runtime implements these operations without significant injection overhead or serialization between concurrent threads. Moreover, successfully injected transfers progress in the background without requiring the runtime polling for progress, or assigning the progress to a thread. While transfer injection is non-blocking by default, a transfer injection could be blocked until resources are available. This back-pressure mechanism provides some throttling mechanism to avoid congesting the interconnect.

By not providing a mechanism of tracking completion of individual transfers or ordering constraints between transfers, the runtime could handle a large number of transfers with minimal overhead.
We structured the communication such that a single space per rank receives the contribution of all producer ranks. Therefore, involving more ranks in the communication does not result in an increase of the overhead of checking the data readiness for consumption. The advantage of such approach manifests at scale.

\section{Numerical Experiments}
\label{sec:exp}

All numerical experiments were performed on the NERSC super computer called Cori, a Cray XC40 system powered by Intel Xeon Phi ``Knights Landing'' (KNL) compute nodes @1.4\,GHz,
68 cores and each with 4 hyper-threads, 96\,GB DDR4 RAM, 16\,GB MCDRAM.
The Cray XC40 nodes are connected using a Dragonfly Aries interconnect. 

Throughout the numerical experiments we use a fixed block size of 64 for the
MATVECs and the LOBPCG eigensolver.
We also use 1 MPI rank per node and OpenMP for on-node parallelism.

\subsection{MPI--OpenMP MATVEC}

\begin{figure}[b!]
\centering
\includegraphics{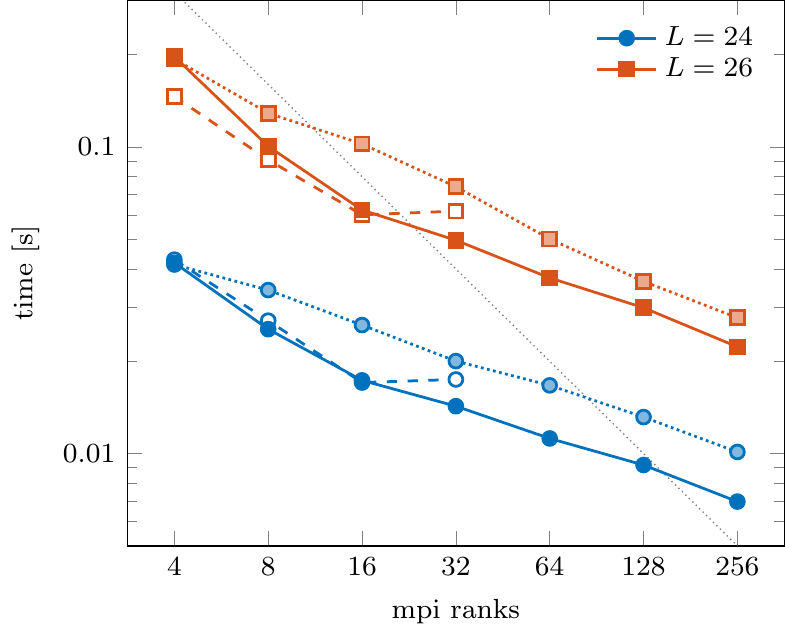}
\caption{\label{fig:matvec-scaling}%
Strong scaling of 1 MATVEC with block size 64.
The dashed lines correspond to the state ordering used in \cite{vbkm2020},
the dotted lines to the METIS ordering, and the solid lines to the METIS
ordering + CSPACER.}
\end{figure}

In a first experiment, we compare the different implementations of the MATVEC.
\Cref{fig:matvec-scaling} shows the strong scaling results for the $L = 24$ and
$L = 26$ MATVECs.
In this figure, the dashed lines correspond to the state ordering of
\cite{vbkm2020} with non-blocking MPI communication and the dotted and solid
lines correspond to the proposed METIS state reordering with non-blocking MPI
and CSPACER communication, respectively.

Due to the large computation and communication imbalance, as shown in
\Cref{fig:matvec-unbalance}, the dashed lines in \Cref{fig:matvec-scaling}
show that the scalability of the MATVEC implementation using the original
ordering of the states stops at 16 ranks.
On the other hand, this figure clearly shows that using the proposed METIS
reordering of the states yields the MATVEC to continue scaling for higher
concurrency.
Although the METIS reordering reduces both the total communication volume and
the maximum communication volume per rank, just changing the ordering of the
states itself does not lead to an improvement of the wall clock time for low
concurrency.
This is because the METIS reordering leads to a more complicated communication
pattern, which could explain the higher overall wall clock time of the dotted
lines in \Cref{fig:matvec-scaling}.
As shown by the solid lines, the extra cost originating from a more complicated
communication pattern can be mitigated with an optimized runtime, for instance using CSPACER.

The quality of the load-balancing of the matrix partitions influences the
scaling behavior because the slowest rank dictates the overall performance.
The graph partitioning is an NP-hard problem and grows in complexity with the
problem size and the number of partitions (rank count).
Later results show that the work distribution, despite being significantly
improved, is not perfectly balanced.
A such, we conjecture that an improved load-balancing will result in a better
scaling behavior for the studied problem.

\subsection{Matrix-Free LOBPCG Eigensolver}

As discussed in \Cref{sec:concurrency}, the MBL problem exhibits multiple levels
of concurrency and requires computing eigenvalues/eigenvectors from different
spectral regions.
Since computing eigenvalues in the middle of the spectrum are the hardest,
we focus on computing eigenvalues around the shift $\sigma = 0$.

In all remaining experiments, we use the matrix-free LOBPCG eigensolver with
the METIS reordering for the MATVEC.
The \texttt{allreduce} operations in the LOBPCG and PCG solver use MPI, in
contrast to the MATVEC for which we compare different communication strategies.
Because most of the MATVECs take place in the preconditioner, we also perform
all PCG iterations in single precision and only the LOBPCG iterations in double
precision.
This mixed precision approach for the MBL problem turns out to have no effect
on the overall eigenvalue accuracy or the total number of LOBPCG iterations,
however, it significantly reduces the wall clock time.

\begin{figure}[h!]
\centering
\includegraphics{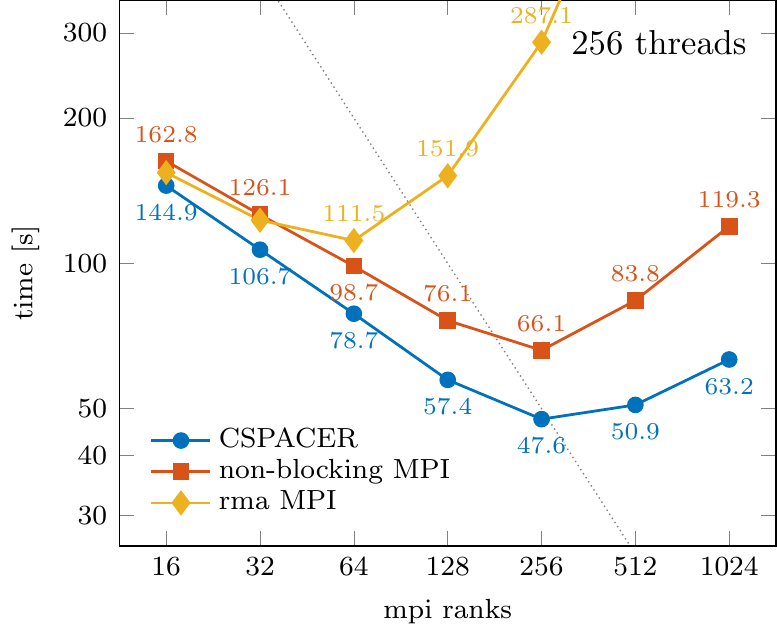}
\caption{\label{fig:L24-scaling-MPI-vs-CSP}%
Strong scaling of 1 $L = 24$ LOBPCG iteration with 5,000 PCG iterations
in single precision.}
\end{figure}

The $L = 24$ strong scaling behavior of the different communication strategies
within the MATVEC are presented in \Cref{fig:L24-scaling-MPI-vs-CSP}.
Note that we only report the timings for 1 $L = 24$ LOBPCG iteration with 5,000
PCG iterations.
In order to converge the eigensolver one needs a few tens of iterations.
From \Cref{fig:L24-scaling-MPI-vs-CSP}, we notice that the CSPACER-variant
outperforms the non-blocking MPI variant and that the difference grows for
increasing concurrency.
On the other hand, one-sided remote memory access (rma) communication
only performs well at low concurrency and is even in that case not competitive
with the CSPACER-variant.
Therefore, due to the poor scalability of the rma MPI-variant, we will not
further consider this type of communication.

\begin{figure}[h!]
\makebox[\textwidth][c]{
\includegraphics{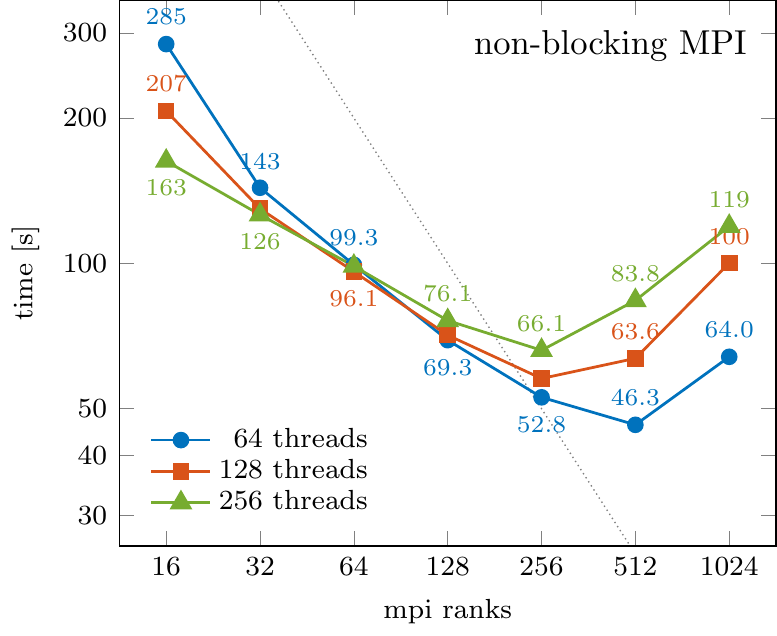}
\quad
\includegraphics{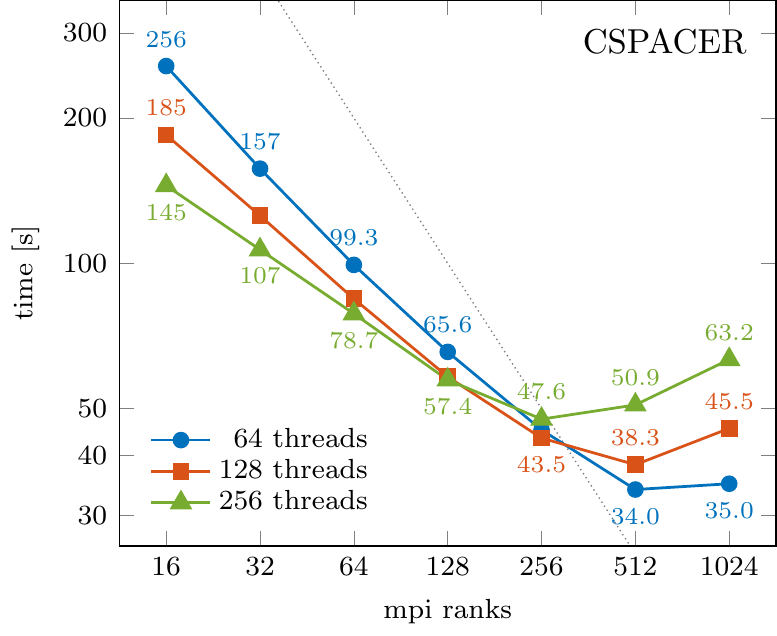}
}
\makebox[\textwidth][c]{
\includegraphics{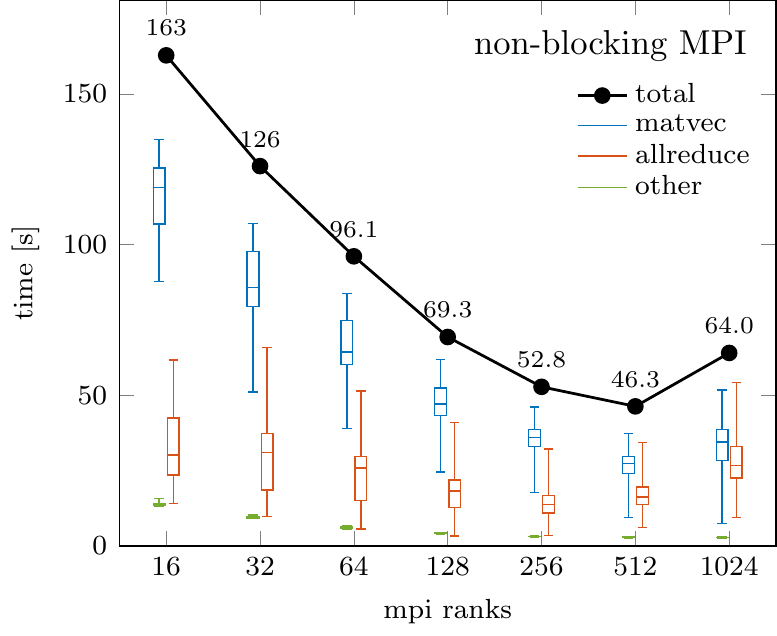}
\quad
\includegraphics{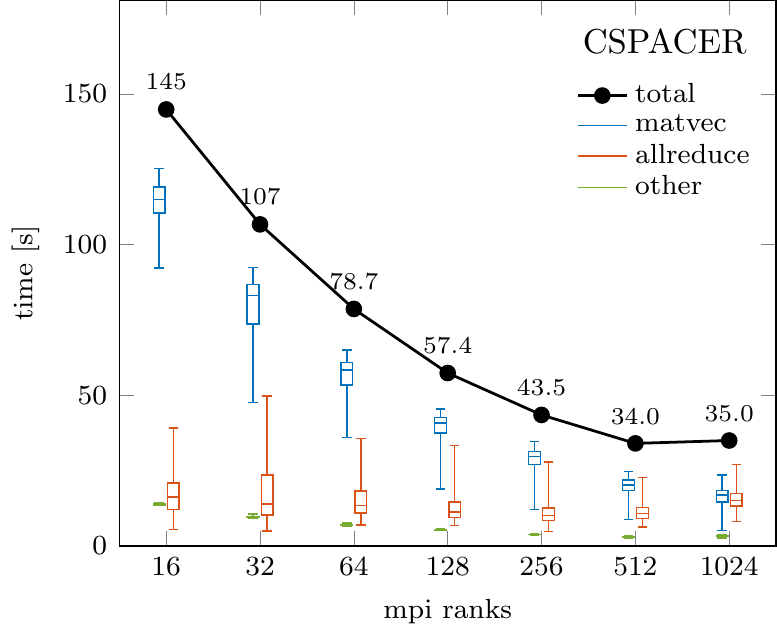}
}
\caption{\label{fig:L24-boxplots}%
Strong scaling of 1 $L = 24$ LOBPCG iteration with 5,000 PCG iterations
in single precision.}
\end{figure}

The upper part of \Cref{fig:L24-boxplots} presents the strong scaling behavior
for the $L = 24$ problem, while using different thread counts per node.
We notice that the need for full thread concurrency diminishes as we scale,
to the extent we start seeing performance degradation at high node concurrency.
This behavior could be attributed to an increased overhead for managing thread
pools, e.g., barrier synchronization for the amount of work assigned to the
threads.
We notice that the CSPACER-variant suffers less from performance degradation
compared to the MPI-variant.
We attribute that to the former using threading more efficiently in injecting
and progressing multiple transfer lanes in the interconnect.
The corresponding speedup factors for the CSPACER-variant compared to
MPI-variant are presented in \Cref{tab:L24}.
We notice that in almost all situations the CSPACER-variant results in a
significant speedup and, as also shown in \Cref{fig:L24-scaling-MPI-vs-CSP,%
fig:L24-boxplots}, the speedup factors increase, in general, for increasing
concurrency.

\begin{table}[b!]
\small\centering
\caption{$L = 24$ speedups for the CSPACER-variant compared to non-blocking MPI
communication.\label{tab:L24}}
\begin{tabularx}{2.85in}{C|S[table-format = 2.1,table-space-text-post = \,\%]S[table-format = 2.1,table-space-text-post = \,\%]S[table-format = 2.1,table-space-text-post = \,\%]}
\toprule
ranks  & \multicolumn{1}{c}{256 threads} &
         \multicolumn{1}{c}{128 threads} &
         \multicolumn{1}{c}{64 threads} \\
\midrule
    16  &   11.0\,\%   &   10.8\,\%   &    9.9\,\% \\
    32  &   15.3\,\%   &    3.3\,\%   &   -9.5\,\% \\
    64  &   20.3\,\%   &   12.1\,\%   &    0.0\,\% \\
   128  &   24.6\,\%   &   18.0\,\%   &    5.4\,\% \\
   256  &   28.0\,\%   &   24.7\,\%   &   14.4\,\% \\
   512  &   39.2\,\%   &   39.8\,\%   &   26.6\,\% \\
  1024  &   47.0\,\%   &   54.5\,\%   &   45.4\,\% \\
\bottomrule
\end{tabularx}
\end{table}

The lower part of \Cref{fig:L24-boxplots} shows the decomposition of the
execution time for the best performing threading configuration for both the
MPI- and CSPACER-variants.
As shown, the MATVEC, blue box, exhibits a significant fraction of the execution time, 
especially at low concurrency. As we strong scale the computation, the
\texttt{allreduce} operations, red box,
start contributing significantly to the execution time. Because the number of elements in the reduction remains 
constant, we expect the overhead of the reduction to increase with the number of nodes.
Instead, we noticed a strong correlation between the variability of the MATVEC and the \texttt{allreduce} phases. 
Given that these two phases are executed consecutively, we conducted an experiment with an extra barrier between them 
and found that the barrier captured most of the variability. We omitted this extra barrier synchronization 
due to its unnecessary overheads in the presented results.

In general, we note that there are multiple sources of performance variability across nodes in our code. 
The first is due to the computational load imbalance originating from the imperfect partitioning;
The second is due to the system noise through the shared interconnect;
The third is due to the communication runtime.
The MATVEC overlap of computation with communication makes it difficult to isolate the impact of 
communication from computation imbalance. 
Having two runtime implementations allow for classifying sources of variability better. 
For instance, the lower variability of the CSPACER-variant compared to the MPI-variant shed some light on
the minimum variability due to the MPI runtime. 
In our experiments, we found the inter-quartile range for the MATVEC variability for MPI compared to CSPACER 
to be $1.5\times$ at low concurrency and reaching $2.7\times$ at 1024 nodes.
We also consider the variability 
of the CSPACER-variant as an upper limit on the computation load imbalance.

\begin{figure}[h!]
\makebox[\textwidth][c]{
\includegraphics{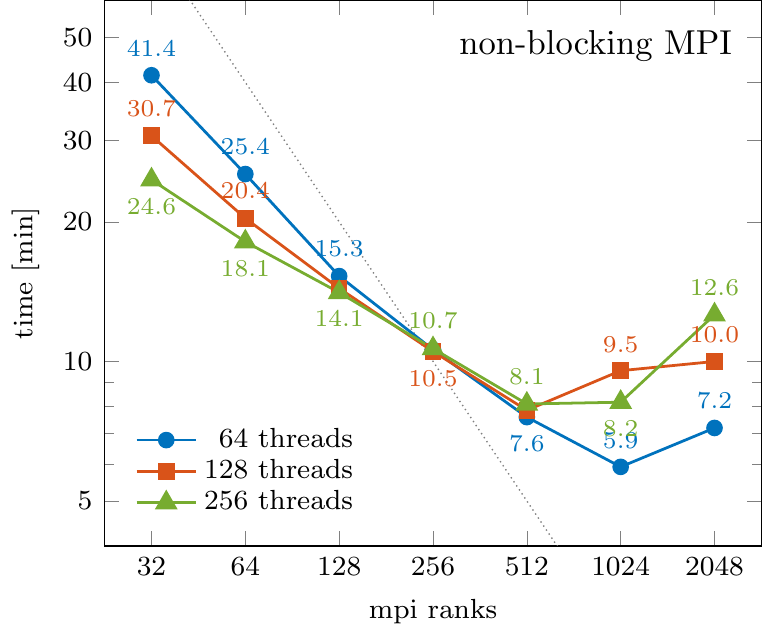}
\quad
\includegraphics{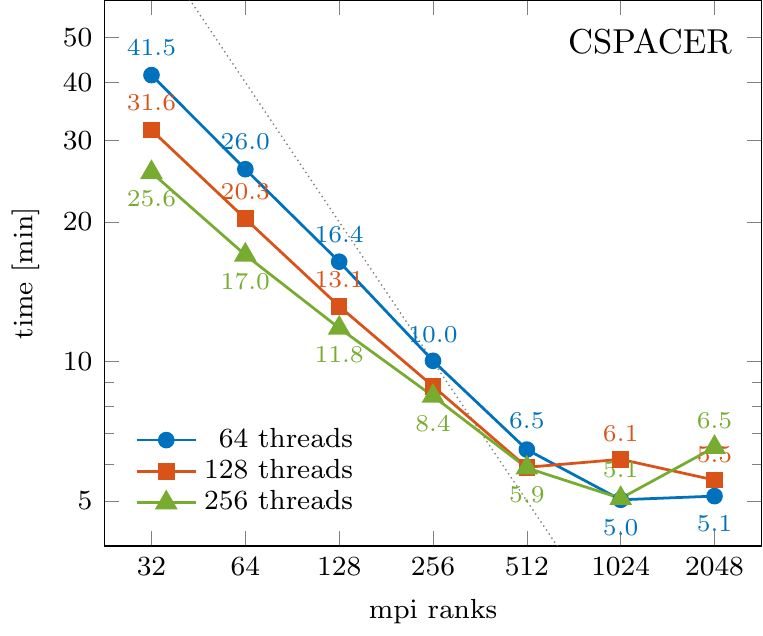}
}
\makebox[\textwidth][c]{
\includegraphics{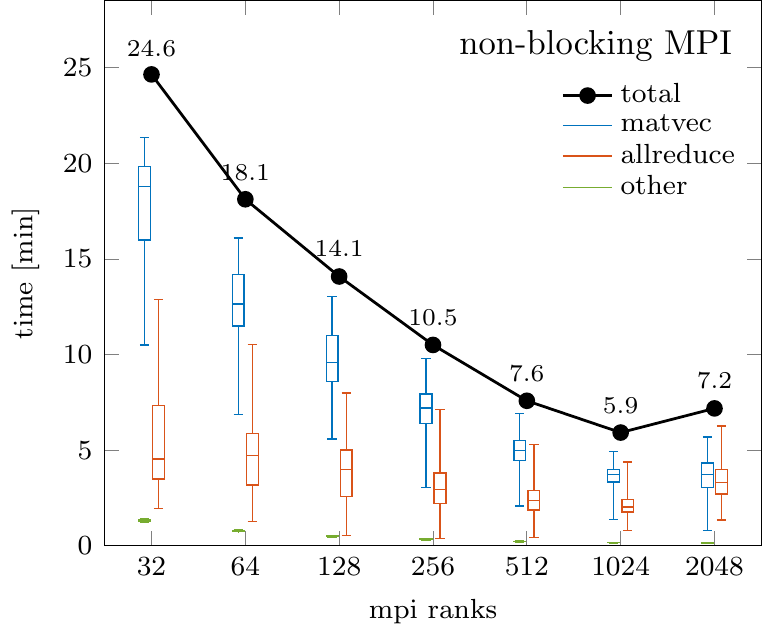}
\quad
\includegraphics{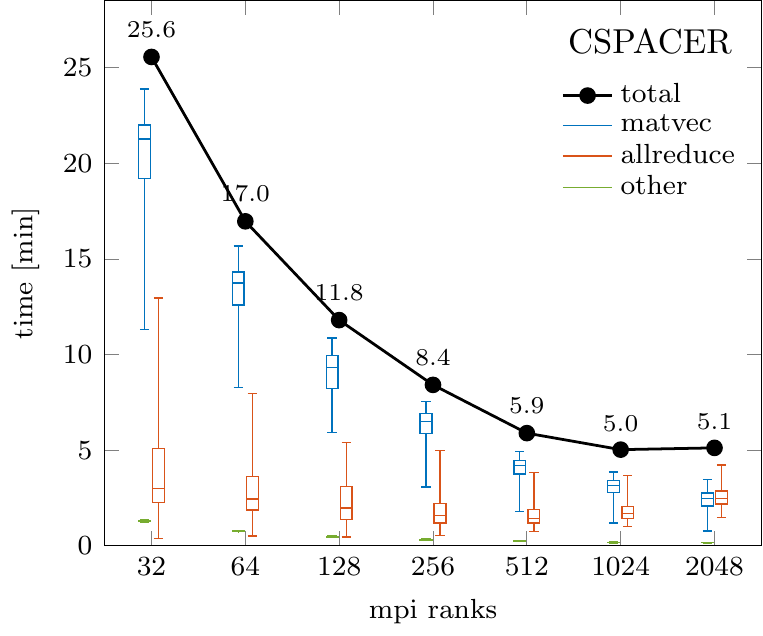}
}
\caption{\label{fig:L26-boxplots}%
Strong scaling of 1 $L = 26$ LOBPCG iteration with 20,000 PCG iterations
in single precision.}
\end{figure}

\begin{table}[b!]
\small\centering
\caption{$L = 26$ speedups for the CSPACER-variant compared to non-blocking MPI
communication.\label{tab:L26}}
\begin{tabularx}{2.85in}{C|S[table-format = 2.1,table-space-text-post = \,\%]S[table-format = 2.1,table-space-text-post = \,\%]S[table-format = 2.1,table-space-text-post = \,\%]}
\toprule
ranks  & \multicolumn{1}{c}{256 threads} &
         \multicolumn{1}{c}{128 threads} &
         \multicolumn{1}{c}{64 threads} \\
\midrule
    32  &   -3.7\,\%   &   -2.9\,\%   &   -0.1\,\% \\
    64  &    6.4\,\%   &    0.2\,\%   &   -2.4\,\% \\
   128  &   16.2\,\%   &    8.5\,\%   &   -7.4\,\% \\
   256  &   21.0\,\%   &   15.8\,\%   &    5.4\,\% \\
   512  &   27.2\,\%   &   24.7\,\%   &   14.9\,\% \\
  1024  &   38.0\,\%   &   35.6\,\%   &   15.1\,\% \\
  2048  &   48.2\,\%   &   44.5\,\%   &   28.7\,\% \\
\bottomrule
\end{tabularx}
\end{table}

\Cref{fig:L26-boxplots} shows the time decomposition and scaling behavior
with various thread concurrency for the $L=26$ problem.
The corresponding speedup factors are given in \Cref{tab:L26}.
While for $L=24$, we noticed performance advantage for the CSPACER-variant across all
concurrency levels, for $L=26$, the performance advantage starts at
128 nodes.
This behavior is somewhat expected because the $L=26$ problem is associated with a larger volume of data movement, making the performance less dependent on the runtime efficiency.
In general, we notice similar trends for the two cases regarding the need to switch to lower thread concurrency as we scale and the correlation of the MATVEC and \texttt{allreduce} variabilities.

We notice that the optimal threading is problem dependent and is likely to change with the underlying systems.  For $L=24$, the advantage for reducing the thread concurrency manifests at 512 nodes, for $L=26$, we need to change the thread-level at 1024 nodes.
Currently, we rely on empirical measurement to identify the best configuration. Leveraging the iterative nature of the algorithm, we could dedicate few iterations for finding the optimal configuration. Ideally, we need to do such change automatically.
We also notice that the optimal thread choice is dependent on the communication runtime. For both presented problem configurations, MPI tends to require switching to lower thread concurrency at lower node count compared to the CSPACER-variant.
The implementation of the latter leverages threads to accelerate the communication progress, which is more advantageous when the number of neighboring ranks increase.

Finally, we compare the overall wall clock time for the $L = 26$ problems
reported in \cite[Table 3]{vbkm2020}.
Using the combination of
    (i) graph partitioning for reducing the communication volume;
   (ii) runtime optimization via CSPACER;
  (iii) performing the MATVECs in the preconditioner only in single precision,
we have been able to significantly increase the scalability of the matrix-free
LOBPCG eigensolver to 1024 nodes.
All these techniques together resulted in a speedup factor of $10\times$ so that
the overall wall clock time for computing eigenvalues/eigenvectors in the middle
of the spectrum (30 LOBPCG iterations with 20,000 PCG iterations as
preconditioner) got reduced from more than 1 day to only 2.5 hours.

\section{Conclusions}
\label{sec:concl}

We have presented several strategies to significantly reduce the computation and
communication imbalance within the matrix-free LOBPCG eigensolver for computing
many eigenvalues and corresponding eigenvectors of large spin Hamiltonians.
Using graph partitioning for reordering the states, both the total communication
volume and the maximum communication volume per rank reduces and enhances the
scalability of the matrix-free eigensolver.
Combining it with communication performance optimization by using CSPACER,
Consistent SPACE Runtime, we have been able to scale the LOBPCG eigensolver up
to 512 and 1024 nodes for $L = 24$ and 26 spins, respectively.
The numerical experiments have illustrated that the proposed techniques of graph
partitioning, runtime optimization, and using mixed precision arithmetic, reduce
the overall wall clock time for the $L = 26$ problem, reported in
\cite{vbkm2020}, by a factor of 10.
Because the MBL study requires solving eigenvalue problems for many instances
of Hamiltonians with random disorder terms, and computing eigenvalues from
different regions of the spectrum, the overall computation can scale to hundreds
of thousands of computational cores.

\section*{Acknowledgements}
This work is partially supported by the U.S. Department of Energy, Office of
Science, Office of Advanced Scientific Computing Research, Scientific Discovery
through Advanced Computing (SciDAC) program and Center for Novel Pathways to
Quantum Coherence in materials, an Energy Frontier Research Center funded by
the US Department of Energy, Director, Office of Science, Office of Basic
Energy Sciences under Contract No. DE-AC0205CH11231.
GDKM was supported by the Department of Defense (DoD) through the National
Defense Science \& Engineering Graduate (NDSEG) Fellowship Program.

This research used resources of the National Energy Research Scientific
Computing Center (NERSC), which is supported by the Office of Science of
the U.S. Department of Energy under Contract No. DE-AC02-05CH11231.

\bibliographystyle{siamplain}
\bibliography{references}

\end{document}